\documentclass[12pt,preprint]{aastex}

\input epsf.sty

\newcommand{\Msun}{M_{\odot}}

\newcommand{\kms}{km\,s$^{-1}$}

\newcommand{\OI}{O\,{\sc i}}

\newcommand{\CI}{C\,{\sc i}}

\newcommand{\MgI}{Mg\,{\sc i}}

\newcommand{\CaII}{Ca\,{\sc ii}}

\newcommand{\FeII}{Fe\,{\sc ii}}
\newcommand{\FeIII}{Fe\,{\sc iii}}
\newcommand{\CoII}{Co\,{\sc ii}}

\newcommand{\NiII}{Ni\,{\sc ii}}
\newcommand{\Fefs}{$^{56}$Fe}
\newcommand{\Cofs}{$^{56}$Co}
\newcommand{\Nifs}{$^{56}$Ni}
\newcommand{\Nife}{$^{58}$Ni}

\def\gsim{\mathrel{\rlap{\lower 4pt \hbox{\hskip 1pt $\sim$}}\raise 1pt \hbox {$>$
}}}
\def\lsim{\mathrel{\rlap{\lower 4pt \hbox{\hskip 1pt $\sim$}}\raise 1pt \hbox {$<$
}}}

\begin{document}

\title{Keck and ESO-VLT View of the Symmetry of the Ejecta of the 
XRF/SN\,2006aj$^0$}

\author{Paolo A.~Mazzali\altaffilmark{1,2,3,4}, 
Ryan~J.~Foley\altaffilmark{5},
Jinsong~Deng\altaffilmark{6,3,4},
Ferdinando~Patat\altaffilmark{7},
Elena~Pian\altaffilmark{2},
Dietrich~Baade\altaffilmark{7},
Joshua~S.~Bloom\altaffilmark{5},
Alexei~V.~Filippenko\altaffilmark{5},
Daniel A. Perley\altaffilmark{5},
Stefano~Valenti\altaffilmark{7,8},
Lifan~Wang\altaffilmark{9,10},
Koji~Kawabata\altaffilmark{11},
Keiichi~Maeda\altaffilmark{12}, and
Ken-ichi~Nomoto\altaffilmark{3,4}
}

\altaffiltext{0}{Based on observations made with ESO Telescopes at the Paranal
	and La Silla Observatories under program 277.D-5039(A).}
\altaffiltext{1}{Max-Planck-Institut f\"ur Astrophysik,
	Karl-Schwarzschild-Str.\ 1, 85748 Garching, Germany.}
\altaffiltext{2}{National Institute for Astrophysics--OATs, Via G.B. Tiepolo, 11,
	34143 Trieste, Italy.}
\altaffiltext{3}{Department of Astronomy, School of Science, 
	University of Tokyo, Bunkyo-ku, Tokyo 113-0033, Japan.}
\altaffiltext{4}{Research Center for the Early Universe, School of Science, 
	University of Tokyo, Bunkyo-ku, Tokyo 113-0033, Japan.}
\altaffiltext{5}{Department of Astronomy, University of California, Berkeley, 
	CA 94720-3411.}
\altaffiltext{6}{National Astronomical Observatories, CAS, 20A Datun Road,
	Chaoyang District, Beijing 100012, China.}
\altaffiltext{7}{European Southern Observatory, Karl-Schwarzschild-Str. 2, 
	85748 Garching, Germany.}
\altaffiltext{8}{Department of Physics, University of Ferrara, 
	Via G. Saragat 1, 44100 Ferrara, Italy.}
\altaffiltext{9}{Physics Department, Texas A\&M University, College Station, TX
	77843-4242.}
\altaffiltext{10}{Purple Mountain Observatory, Chinese Academy of
	Sciences, 2 Beijing Xi Lu, Nanjing, Jiangsu 210008, China.}
\altaffiltext{11}{Hiroshima Astrophysical Science Center,
        Hiroshima University, Hiroshima 739-8526, Japan.}
\altaffiltext{12}{Department of Earth Science and Astronomy, Graduate School of
	Arts and Science, University of Tokyo, 3-8-1 Komaba, Meguro-ku, Tokyo
	153-8892, Japan.}

\begin{abstract} 
Nebular-phase spectra of SN\,2006aj, which was discovered in coincidence with X-ray
flash 060218, were obtained with Keck in 2006 July and the Very Large Telescope in
2006 September.  At the latter epoch spectropolarimetry was also attempted,
yielding an upper limit of $\sim$2\% for the polarization. The spectra show strong
emission lines of [\OI] and \MgI], as expected from a Type Ic supernova, but weak
\CaII\ lines. The [\FeII] lines that were strong in the spectra of SN\,1998bw are
much weaker in SN\,2006aj, consistent with the lower luminosity of this SN. The
outer velocity of the line-emitting ejecta is $\sim 8000$\,\kms\ in July and $\sim
7400$\,\kms\ in September, consistent with the relatively low kinetic energy of
expansion of SN\,2006aj. All emission lines have similar width, and the profiles
are symmetric, indicating that no major asymmetries are present in the ejecta at
the velocities sampled by the nebular lines ($v < 8000$\,\kms), except  perhaps in
the innermost part. The spectra were modelled with a non-LTE code.  The mass of
\Nifs\ required to power the emission spectrum is $\sim 0.20\,\Msun$, in excellent
agreement with the results of early light curve modelling. The oxygen mass is $\sim
1.5\,\Msun$, again much less than in SN\,1998bw but larger by $\sim 0.7\,\Msun$
than the value derived from the early-time modelling.  The total ejected mass is
$\sim 2\,\Msun$ below 8000\,\kms.  This confirms that SN\,2006aj was only slightly
more massive and energetic than the prototypical Type Ic SN\,1994I, but also
indicates the presence of a dense inner core, containing $\sim 1\,\Msun$\ of mostly
oxygen and carbon. The presence of such a core is inferred for all broad-lined
SNe~Ic. This core may have the form of an equatorial oxygen-dominated region, 
but it is too deep to affect the early light curve and too small to affect the 
late polarization spectrum. 
\end{abstract}


\section{Introduction}

The connection between some gamma-ray bursts (GRBs) and supernovae (SNe) is now
well established (see \citealt{wb06} for a recent review). The first observational
clue for a connection came with the discovery of a bright SN coincident in time and
space with GRB\,980425 \citep{gal98,pian00}. The SN was eventually classified as a
Type Ic \cite[for a definition of the class, see][]{fil97}, indicating an origin
from a massive star that lost both the H and He envelopes prior to core collapse. 
SN\,1998bw was an exceptional SN~Ic in being very luminous and showing a
broad-lined spectrum, indicative of a large kinetic energy of expansion. Such
objects have been called hypernovae \citep{iwa98}  or broad-lined SNe~Ic
\citep[SNe~Ic-BL; see][]{wb06}.  Modelling confirmed the large kinetic energy,
$\sim (10-30) \times 10^{51}$\,erg, and showed that the ejected mass  was also
quite large \citep[$\sim 10\,\Msun$,][]{iwa98}, pointing to a  progenitor star with
zero-age main sequence (ZAMS) mass $\sim 40\,\Msun$.  

Later, other nearby long-duration soft-spectrum GRBs, in the same high-energy class
as GRB 980425, were also seen to accompany hypernovae/SNe~Ic-BL, confirming the
connection \citep{sta03,mal04}. The SNe had properties very similar to those of
SN\,1998bw, although the properties of the GRBs were quite different
\citep{maz03,maz06a}. In a popular model, both the SN and the GRB are produced when
the rapidly rotating core of a massive star collapses to form a black hole
\citep{McFW99}. The coincidence of a very aspherical event, the GRB, and a SN
motivates the search for signatures of asphericity in the SN as well. 

In the case of SN\,1998bw, evidence of asphericity came from the nebular spectra. 
These showed strong emission lines of both [\OI]  $\lambda\lambda$6300, 6363 and
[\FeII] (a blend near 5200\,\AA), as could be expected from a bright SN~Ic, but
with the peculiarity that even after accounting for blending the [\FeII] lines were
broader than the [\OI] line, indicating a higher expansion velocity of Fe than of
O. This was explained in the context of an axisymmetric explosion, where the Fe
produced by the decay of \Nifs\ is ejected at high velocities near the direction of
the GRB jet while oxygen, which is mostly not a product of the explosion but rather
left over from the progenitor, is ejected at lower velocities nearer the equator
\citep{maz01}. \cite{mae02} supported this scenario with two-dimensional explosion
models and three-dimensional synthetic nebular spectra, and suggested that our
viewing angle was $\sim 15-30^{\circ}$ from the jet axis \citep[see also][]{mae06}.
This may partly be responsible for the weakness of GRB980425 \citep{err05}. 

In the case of SN\,2003dh/GRB030329, polarization levels of $\sim2$\% were observed
in the first 2--3 days after the GRB, but they were probably measurements of the
GRB afterglow \citep[][and references therein]{kaw03,gre03}, dominated by the
synchrotron process which imposes partial polarization. Later polarization
measurements, which should not have been affected by the afterglow, suggested that
the polarization was small ($P<1$\%), although the alignment of the polarization
angle suggests that the polarization is non-zero \citep{kaw03}. This could be
consistent with a jet-like explosion viewed very close to the jet axis, as the
strength of GRB\,030329 may suggest.  

A later confirmation that the ejecta of bright SNe~Ic can be very aspherical
came from the detection of the double-peaked emission profile of the [\OI]
$\lambda\lambda$6300, 6363 line in SN\,2003jd \citep{maz05}. 
The profile could be reproduced
assuming that we observed an explosion similar to that of SN\,1998bw from close
to the equatorial plane. A GRB may or may not \citep{sod06a} have been
produced, and if it was it may not have been detected because it was not
pointing toward the Earth.

Asphericity may indeed be a common feature of core-collapse SNe \citep{leo06}, and
it is probably stronger in the carbon-oxygen cores, since envelope-stripped SNe
show a higher degree of polarization \citep{wang96,fl04,lf05}. Spectropolarimetry 
of the non-GRB, broad-lined SN\,Ic 2002ap \citep{kaw02,leo02,wang02} revealed
polarization of up to $\sim2$\% in the \OI\ $\lambda$7774 line and in the \CaII\ IR
triplet at early times, suggesting a deviation from sphericity of $\sim 10-20$\%. 

X-ray flashes (XRFs) are a lower-energy and softer subclass of long-duration GRBs
\citep{heise01}. Previous work had indicated a connection between XRFs and SNe
\citep{fyn04,tom04,sod05} but high-quality spectra of the SNe had not been obtained
until the Type Ic SN\,2006aj was discovered coincident with XRF\,060218, at a
redshift $z = 0.03342$ \citep{pian06}. Although SN\,2006aj showed a broad-lined
spectrum, this was not as broad as those of SNe\,1998bw and 2003dh. Therefore,
SN\,2006aj could not, {\em prima facie}, be considered an extremely energetic SN
from early optical spectroscopy and photometry. Radio observations confirmed this
\citep{sod06b}.

SN\,2006aj was modelled \citep{maz06b} to have synthesized only $\sim 0.2\, \Msun$
of \Nifs, and to have ejecta only slightly more massive ($\sim 2\, \Msun$) and
energetic ($E \approx 2 \times 10^{51}$\,erg) than ``normal" SNe~Ic.\footnote{We
use the term "normal" to refer to SNe like SN~1994I \citep[][and references
therein]{sau06}, i.e. SNe with explosion energy $\approx 10^{51}$\,erg, although
it is by no means clear that this can be called a normal event.} These parameters
are more appropriate for the collapse of a star of $\sim 20\, \Msun$ to a neutron
star. \cite{maz06b} suggested that magnetic activity on the surface of the nascent
neutron star increased the energy of the explosion and caused the XRF. In this
case, the explosion may not be highly aspherical. The absence of a jet break and
the behavior of the radio flux led \citet{sod06b} to estimate a ``jet" opening
angle of $\gsim 50^\circ$. 

Large polarization ($\sim4$\% at 3--5 days) was observed in SN\,2006aj
\citep{gor06}. As in the case of SN\,2003dh, this may have been due to the
relativistic jet and the XRF afterglow synchrotron continuum, but as the data have
a later epoch than those of SN\,2003dh \citep{kaw03}, and the afterglow was not as
strong as in the case of SN\,2003dh, an aspherical fast outflow from the outermost
SN ejecta may also have been detected. At later epochs, \citet{gor06} find that the
polarization level dropped to $\sim1.4$\%, which they attribute to the host
galaxy. 

Obviously, it is important to verify these suggestions using nebular spectroscopy.
We observed SN\,2006aj with the Keck\,I telescope and the Very Large Telescope
(VLT) when it became visible again after solar occultation. In the following, we
discuss the observations (\S 2), present the observational results (\S 3) and the
models of the nebular spectra (\S 4), and discuss our findings (\S 5).

\section{Observations and Spectral Analysis}


On 2006 July 26.61 (UT dates are used throughout this paper), SN\,2006aj was 
observed with the 10-m Keck\,I telescope on Mauna Kea, Hawaii, equipped
with the Low Resolution Imaging Spectrograph \citep[LRIS;][]{Oke95} with a D560
dichroic to separate the blue and red light paths. The seeing was about
$0.8''$, and a long slit of width $1''$ was aligned along a position angle 
of $-82^\circ$. Although the parallactic angle (Filippenko 1982) at that 
time was $-52^\circ$, the atmospheric dispersion was not very significant,
given the relatively low airmass (1.3). The blue-side data were
taken with the 600/4000 grism, giving a spectral resolution of 3.5\,\AA, 
while the red-side data were taken with the 400/8500 grating, giving a 
spectral resolution of 6.4\,\AA.  Two 600\,s exposures were obtained.  
Standard CCD processing was accomplished with IRAF\footnote{IRAF is 
distributed by the National Optical Astronomy Observatories, which are 
operated by the Association of Universities for Research in Astronomy, 
under contract with the National Science Foundation.}. The data were 
extracted using the optimal algorithm of \citet{Horne86}. Wavelength 
calibration was obtained from internal arc lamps and slight adjustments 
were derived from night-sky lines.  The spectrophotometric standard 
stars BD+284211 and BD+174708 (observed at the parallactic angle), as 
well as our own IDL routines, were used for flux calibration and 
telluric absorption removal \citep{Wade88, Matheson00}.


On 2006 September 19.33 and 20.36, SN2006aj was observed with the 8-m VLT Kueyen of
the European Southern Observatory (ESO) at Cerro Paranal, equipped with the FOcal
Reducer Spectrograph \citep[FORS1,][]{app98} in polarimetric mode. The observations
were performed with the 300V grism and a $1.1''$ wide slit, aligned along the
parallactic angle, giving a dispersion of $\sim$2.6\,\AA\/ pixel$^{-1}$ and a
resolution of $\sim$11\,\AA\ at 5500\,\AA\ (full width at half-maximum  intensity).
The wavelength coverage achieved with this setup is 3300--8600\,\AA. At both epochs
the SN was observed at four half-wave plate position angles ($0^\circ$,
$22.5^\circ$, $45^\circ$, and  $67.5^\circ$), for a total integration time of
7200\,s. The seeing was $1.2''$ on September 19 and $1.5''$ on September 20. 

The data were bias-subtracted, flattened, and wavelength-calibrated using standard
tasks within IRAF. Stokes parameters, linear polarization degree, and position
angle were computed using specific routines written by the authors.  Finally,
polarization bias correction and error estimates were performed following the
prescriptions described by \citet{patat06a}, while the half-wave plate zeropoint
angle chromatism was corrected using FORS1 tabulated data\footnote{See {\tt
http://www.eso.org/instruments/fors/inst/pola.html}}.  In order to reduce the
noise, the two data sets have been combined and the final Stokes parameters
rebinned to 75\,\AA\ bins (29 pixels). The signal-to-noise ratio (S/N) in the
binned flux spectrum is $\sim$100 in the [\OI] $\lambda\lambda$6300, 6363 region,
corresponding to an expected root-mean square uncertainty in the  polarization
degree of $\sim0.7$\%.

Photometric calibration of the spectra was achieved by observing spectrophotometric
standard stars with full polarimetric optics inserted. Instrumental polarization
and position angle offset were checked by observing polarized and unpolarized
standard stars, within the FORS1 calibration plan.

The total-flux spectrum was obtained combining all exposures, giving S/N $\approx
30$ in the continuum at 6000\,\AA. At this advanced nebular phase, the only
relevant SN signal is expected to be associated with the nebular emission lines. 
Therefore, we attributed all observed continuum to the host galaxy. The continuum
was evaluated in four spectral regions (near 4400\,\AA, i.e., immediately blueward
of the \MgI] line; at 6200\,\AA\ and 7020\,\AA, i.e., on either side of the [\OI]
line; and at 8170\,\AA), and was then interpolated at all wavelengths of our
spectral range with a spline and subtracted, along with the host-galaxy emission
lines, from the observed spectra. We thus obtained the pure SN emission line
spectrum.  

In order to compute line intensities the spectral flux was converted to physical
units. The SN spectrum is superposed on the host-galaxy continuum. Galaxy 
photometry, however, includes narrow host-galaxy emission lines 
\citep{cool06,mod06,pian06,soll06}. The flux level of the SN spectra is however
best established by comparing to the galaxy continuum. Therefore, first we
estimated from the observed spectra the fractional contribution of the host-galaxy
emission lines to the galaxy photometry in each filter, and hence determined the
photometric level of the galaxy spectral continuum. This allowed us to evaluate the
flux level of the observed and subtracted SN spectra, and ultimately to estimate
the intensities of the SN nebular emission lines.  As the host galaxy of SN\,2006aj
is very small and compact, the fraction of galaxy and SN light possibly not
included in the slit during spectroscopy should be similar, ensuring that our
method does not substantially underestimate the galaxy contribution. Figure 1 shows
the observed VLT spectrum, the fiducial galaxy spectrum, the observed spectrum
recalibrated according to the galaxy photometry, as explained above, and the
subtracted SN spectrum, calibrated and cleaned of cosmic rays.

\section{Results}

Close inspection of the VLT-FORS1 polarization data shows no detection of linear
polarization, either in the continuum or in the most prominent emission line, [\OI]
$\lambda\lambda$6300, 6363; only a conservative 3$\sigma$ upper limit of 2\% across
the whole wavelength range could be set. Considering that a polarization  level of
$P \approx 1.4$\% may be attributed to the host galaxy \citep{gor06},  and that
interstellar polarization may contribute a similar amount $[P_{IS,max} =  0.9
\times E(B-V)]$, this is consistent with unpolarized SN flux, and with a global 
ellipticity of the inner SN ejecta (i.e. at $v < 8000$\,\kms) of no more than 10\%
\citep{hoef91}. This is not expected to be observable in the profiles of the
nebular emission lines. 

We derived SN magnitudes $B\approx 22.8$, $V\approx 21.6$, $R\approx 21.0$, 
and $I\approx 20.9$ for the Keck data (rest-frame epoch 153 days),  
and $B\approx 23.4$, $V\approx 22.6$, and $R\approx 21.8$ for the VLT data 
(rest-frame epoch 206 days). We assign an uncertainty of $\sim 0.1$ mag 
in the $R$, $V$, and $I$ bands, and 0.2 mag in the $B$ band. The average
light curve decline rate in this period is then $\sim 1.7\pm 0.7$, $1.5\pm 
0.4$, and $1.2\pm 0.3$\,mag\,(100 days)$^{-1}$ in the $V$, $R$, and $I$ 
bands, respectively.  This is consistent with expectation for a late-time 
light curve powered by gamma rays from $^{56}$Co decay in relatively 
transparent SN ejecta and is also consistent with other SNe~Ibc such as 
SN\,1998bw \citep{patat01} and SN\,2002ap \citep{fol03,tom06}.

The Keck spectrum, obtained on 2006 July 26, has a rest-frame epoch of 153 days.
Figure 2 shows a comparison with the SN\,1998bw spectrum obtained on 1998
September 12, at an epoch of 139 days after the explosion, which is assumed to
have occurred on 25 April 1998, in coincidence with GRB\,980425 \citep{pian00}.
The two spectra are very similar, but obviously the [\OI] line is stronger
relative to all other lines in SN\,2006aj. 

The VLT spectrum, obtained on 2006 Sept 19--20, has a rest-frame epoch of 206 days.
It is shown in Figure 3 compared to the spectrum of SN\,1998bw obtained on 1998
November 26, at a rest-frame epoch of 216 days after the explosion.  Again, [\OI]
$\lambda\lambda$6300, 6363 is by far the strongest emission line, and is stronger
in SN\,2006aj than in SN\,1998bw relative to the other lines, indicating that
oxygen dominates the composition of the ejecta. The next strongest lines are near
4600\,\AA, corresponding to \MgI] $\lambda$4571\,\AA, and a line near 7400\,\AA.
This is too red to be \CaII] $\lambda\lambda$7291,7324, and [\NiII] $\lambda$7380 is
the best candidate \citep{mae07}.  The wavelength of the emission near 8700\,\AA\
in the Keck spectrum suggests that the \CaII\ IR triplet is not the only
contributor, and that [\CI] $\lambda$8727 is also strong.  The [\FeII] lines are
relatively weak, but still form an emission feature near 5300\,\AA. The strongest
lines in that complex have rest wavelengths 5159, 5262, 5273, and 5334\,\AA.
[\FeIII] lines are very weak, as in all SN~Ic spectra, indicating low temperature
and significant clumping.

The profile of the emission lines in the nebular spectrum of SN\,2006aj is not very
sharp, indicating that the nebula is to a good approximation spherically
symmetric.  Figure 4 shows a comparison of the [\OI] $\lambda\lambda$6300, 6363
line of SNe 1998bw, 2006aj, and 2003jd. The line in SN\,2006aj is broader than the
other two by $\sim 2000$\,\kms, although the ejecta velicity of SN\,2006aj was
significantly lower than that of SN\,1998bw. The line of SN\,1998bw is quite sharp,
which was interpreted as a disk-like distribution of slow-moving oxygen viewed from
a near-polar direction \citep{mae02}. In SN\,2003jd the line has a width similar to
that of SN\,1998bw, but the double-peak profile suggests that we are viewing the
oxygen-rich disk from near its plane \citep{maz06b}.

\section{Models}

We modelled the nebular spectra using a non-LTE code \citep{maz01} based on the
approximations discussed in \cite{axe80} and \cite{RLL92}. The deposition of the
gamma rays emitted in the decay of \Cofs\ to \Fefs\ is computed using a gray
opacity, while the positrons that are also produced are assumed to deposit their
energy {\em in situ}. Collisional heating by the fast  particles produced by the
deposition of the gamma rays and the positrons is computed, and is balanced by
cooling via line emission. The emission lines depend on the composition; in the
case of a SN~Ic, oxygen dominates the cooling. 

For both spectra the spherical approximation seems reasonable, and so we use the
code in its simplest version, assuming that the emitting volume is spherical and
homogeneous. This allows us to test the basic geometric properties of the ejecta
of SN\,2006aj by verifying any deviation of the nebular lines from a parabolic
profile. We adopted a distance of 140\,Mpc and a reddening correction of
$E_{B-V}$ = 0.13 mag \citep{pian06}, which takes into account both Galactic and
intrinsic absorption. 

The main values of the fits are shown in Table 1. The outer velocity of the
model for the Keck spectrum is 8000\,\kms, while for the later VLT spectrum
this value is 7400\,\kms. A slight decrease of the outer velocity may be
expected as the expansion makes the density decrease with time, such that outer
regions may progressively become too thin for collisional excitation processes
to work efficiently. 

The Keck spectrum (Figure 5) requires a \Nifs\ mass of $0.20 \pm 0.01\,\Msun$, in
excellent agreement with the early light-curve model \citep{maz06b}, and an oxygen
mass of $1.50 \pm 0.15\, \Msun$. Uncertainties are estimated from parameter
combinations that yield acceptable fits. The line near 7400\,\AA\ is mostly due to
[\NiII] $\lambda$7380. Given the late epoch of the spectra, stable \Nife\ must be
responsible for the emission, which can be reproduced assuming a \Nife\ mass of
$0.02\, \Msun$. The consequences of this on the properties of the progenitor star
are discussed in an accompanying paper \citep{mae07}. Other lines contributing to
the emission are [\FeII] $\lambda\lambda$7155, 7453 and [\CoII] $\lambda$7541.  The
\CaII] $\lambda\lambda$7291, 7324 lines are very weak. Similarly, the emission near
8600\,\AA\ is mostly due to [\CI] $\lambda$8727, while the \CaII\ IR triplet is
very weak. The line can be reproduced only if we assume the presence of $\sim 0.3\,
\Msun$ of carbon. This is an interesting result, as in the early phase no carbon
was visible in the spectrum \citep{maz06b}. The low calcium abundance is also
interesting, as it may be a clue of the nucleosynthesis in the progenitor and in
the explosion. The Ca/O fraction is $\sim 10^{-4}$ by mass, whereas the
corresponding value for SN\,1998bw was $\sim 10^{-2}$. The total mass enclosed
within the outer velocity of 8000\,\kms\ is $2.07 \pm 0.20\, \Msun$.


The \Nifs\ mass estimated for the VLT spectrum is $0.19 \pm 0.01 \, \Msun$ and the
oxygen mass is $1.42 \pm 0.15 \,\Msun$ (Figure 6). The total mass enclosed within
the outer velocity of 7400\,\kms\ is $1.94 \pm 0.25\, \Msun$. The slight reduction
in ejected mass derived from the later VLT spectrum with respect to the earlier
Keck spectrum is a consequence of the slightly narrower emission lines, and is
consistent with the model used to fit the early light curve and spectra 
\citep{maz06b}. 
However, while the mass of \Nifs\ derived from the nebular models ($\approx 0.20 
\Msun$) is in excellent agreement with the value estimated from the early
light-curve evolution, the ejected mass, and in particular the masses of oxygen and
carbon, are larger. 

The last early-time spectrum, obtained on 2006 March 10 \citep{pian06}, was
modelled for a photospheric velocity of 10,000\,\kms, so there is no overlap
between the region sampled by the early data (above 10,000\,\kms) and that sampled
by the nebular spectra (below 8000\,\kms), except that the early light curve
depends on the density of the innermost regions. The ejected mass predicted below
8000\,\kms\ by the model used by \cite{maz06b} is $\sim 1 \, \Msun$, so the nebular
results suggest the presence of a dense inner region containing an additional $\sim
1 \,\Msun$ mostly composed of oxygen ($\sim 0.7\, \Msun$) and carbon ($\sim 0.3\,
\Msun$), signs of which were not visible in the early-time spectra. The presence of
an inner high-density core is typical of other SNe~Ic (e.g., SN 1998bw, SN 2002ap),
as discussed in the next section.

The only deviation from a parabolic profile may be seen in the core of the [\OI]
line. This may be due to the assumption of a constant density, or to the
presence of an oxygen-dominated high-density inner core.  The profile suggests
that the oxygen abundance is dropping at velocities between 2000 and 4000\,\kms,
but is higher again at the lowest velocities $(v \leq 2000$\, \kms). Between 2000
and 4000\,\kms\ intermediate-mass elements may dominate, but the high flux at
the lowest velocities may indicate the presence of a disk-like oxygen-rich
region. Further data and more detailed modelling will clarify this.

\section{Discussion} 

The nebular spectra of SN\,2006aj are dominated by strong [\OI]
$\lambda\lambda$6300, 6363 emission, as expected for a SN~Ic. The masses derived
from the modelling are small, in general agreement with the values obtained from
the early-phase modelling, but indicate the presence of a central low-velocity zone
containing $\sim 1\, \Msun$ of mostly carbon and oxygen. The abundance of calcium
is also very low in the nebula.  The presence of a dense, oxygen-rich core has been
deduced from the nebular spectra \citep[SN\,1998bw,][]{maz01}, or via the analysis
of the light curves \citep{mae03} in all well-observed broad-lined SNe~Ic, and it
is a strong indication that the explosion was not spherically symmetric. Only if
the mass of this zone is large can an effect on the early light curve be
appreciated \citep[e.g., SN\,1997ef,][]{mae03}. Unfortunately, in the case of
SN\,2006aj spectroscopic observations in the early phase extended only to 2006
March 10 (20 days after the explosion), and the evolution of the light curve at
epochs of 2--3 months, when the linear decline is a signature of the inner dense
zone \citep{mae03}, was missed. Nevertheless, the power of a combined approach of
studying both the early phase and the late phase is highlighted by the present
results.

The nebular lines have rather symmetric profiles. In particular, the [\OI] line is
not as sharp as in SN\,1998bw, indicating a low degree of asymmetry. Early radio
observations suggested that any aspherical outflow was characterized by a large
opening angle \citep{sod06b}, and our results confirm that the ejecta of SN\,2006aj
look more like those of ordinary SNe~Ic \citep[e.g., SN\,1994I,][]{fil95,sau06}
than those of SN\,1998bw. Early-phase modelling suggested that only a small
fraction of the \Nifs\ synthesised by the SN was ejected at high velocities, and
our nebular models confirm that most of the \Nifs\ was ejected at low velocities. 

The only possible deviation from sphericity may be in the deepest regions, as
indicated by the low-velocity enhancement of the [\OI] line. This may indicate an
oxygen-rich disk viewed nearly face-on, as in SN\,1998bw but of much smaller size,
confirming that all SNe~Ic are probably intrinsically aspherical in their cores
\citep{lf05,leo06}. Given the small size of this region, and limited quality of our
polarization data, we do not expect that we can detect polarization from that
structure. Additionally, at the late phases of these data, polarization is more
sensitive to the distribution of \Nifs\ than to global asymmetries of the ejecta.
The lack of a detection of polarization may therefore also indicate that the \Nifs\
distribution in the inner ejecta was not very aspherical.

Our results confirm that the mass ejected in the SN explosion was small, and thus
lend credence to our earlier conclusion that the progenitor star had a ZAMS mass of
$\sim 20\, \Msun$, and that probably the remnant was a neutron star (Mazzali et al.
2006b; see also Maeda et al. 2007).
They may also shed some light on the magnetar phenomenon. If the birth of a
magnetar was at the origin of the XRF, this magnetic activity did not lead to gross
asphericity in the SN ejecta or to the synthesis of much \Nifs\ at high velocities.
Probably, it did not lead to a jet at all, but rather to broad polar outflows. This
should be useful not only for further modelling, but also to understand better the
nature of XRFs.  As mentioned by \citet{pian06}, XRFs may be intrinsically weaker
and more spherical versions of the classical, highly collimated  cosmological
GRBs. 

Continued monitoring of SN\,2006aj is planned at both Keck and VLT.
When more data are collected, we will study in detail the evolution of the
spectra and the light curve in the late phase. 

We thank the ESO Director for allocating time to this program. We also thank 
the night astronomers of the Paranal Science Operations Team, the staff 
of the Keck Observatory, and Dan Kocevski for help with the observations.
This work was partially supported by the Italian Ministry for University and 
Research (MIUR) under COFIN 2004 ``The Physics of the Explosion of Massive 
Stars.'' This work was partially conducted by the groups of J.S.B. and A.V.F.
under US Department of Energy SciDAC grant DE-FC02-06ER41453. A.V.F. is
also grateful for the support of NSF grant AST-0607485.





\begin{figure}
\plotone{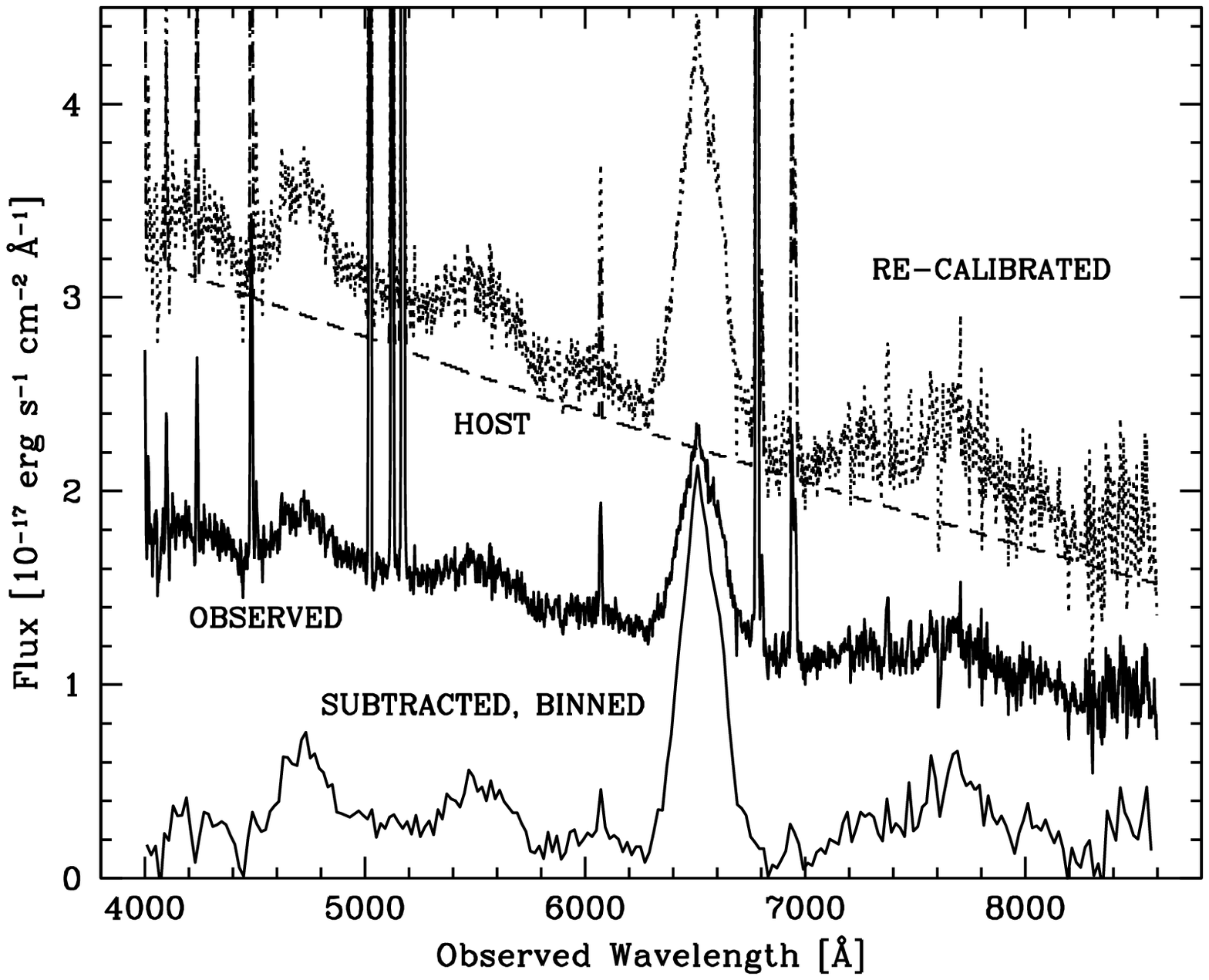}
\figcaption[]{VLT-FORS1 spectrum of SN\,2006aj on 2006 September 19--20.   
The middle solid curve shows the observed spectrum.  The upper thin dotted curve 
is the observed spectrum re-calibrated according to the host-galaxy photometry 
(see Section 2).  The host-galaxy spectrum is shown as a dashed curve.  
The spectrum of the SN, obtained by subtracting the galaxy spectrum (including the
narrow emission lines) from the re-calibrated spectrum and binned using a 20~\AA\ 
boxcar, is the lower solid curve.}
\end{figure}


\begin{figure}
\plotone{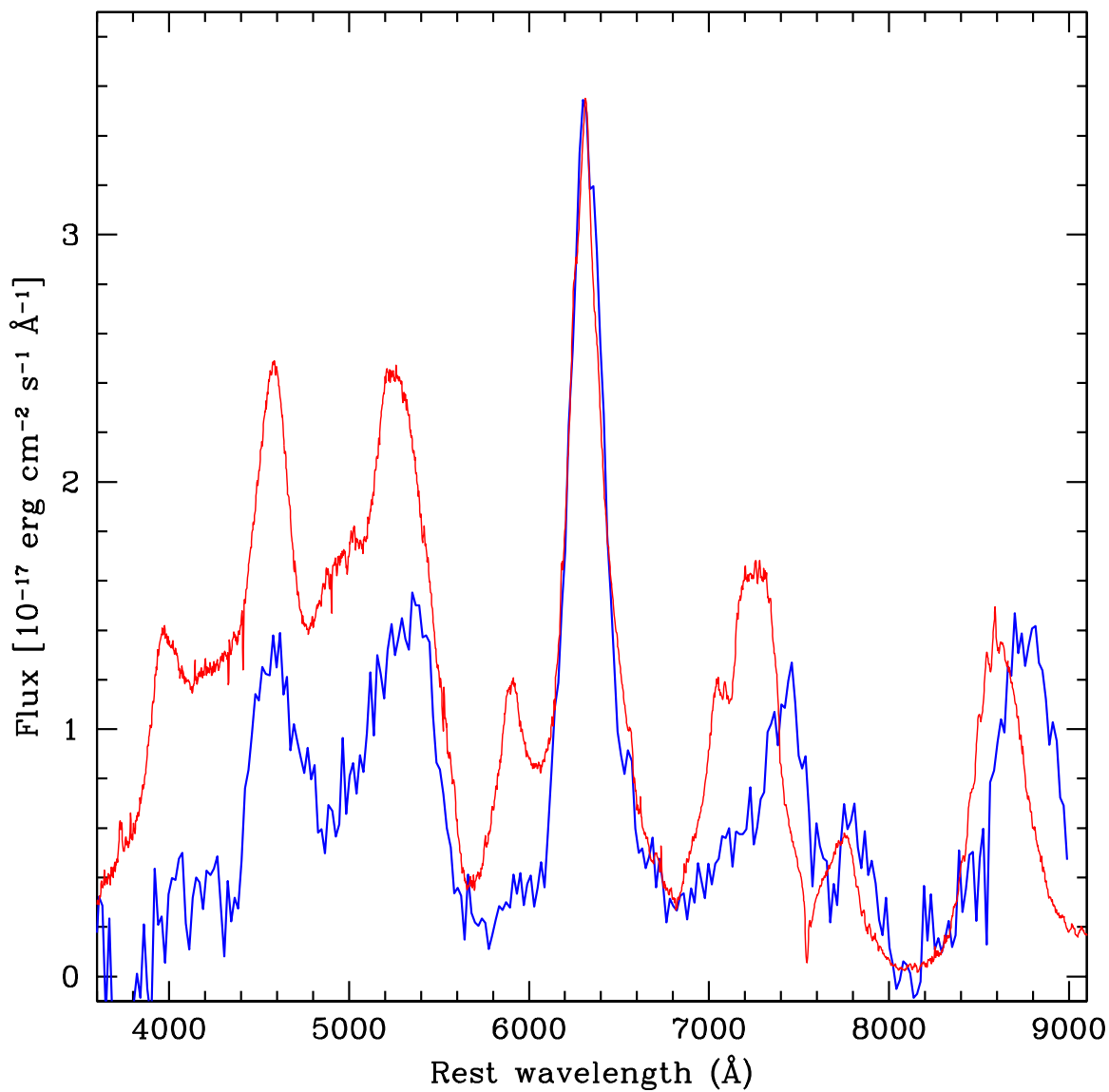}
\figcaption[]{The spectrum of SN\,2006aj obtained at the Keck Observatory on 
2006 July 26, corresponding to a rest-frame epoch of 153 days after XRF\,060218 
(solid/blue line), calibrated and binned with a 25~\AA\ boxcar, compared to the spectrum 
of SN\,1998bw spectrum obtained on 1998 September 12 (Patat et al. 2001), 
at an epoch of 139 days after the explosion, which is assumed to have occurred 
on 25 April 1998, in coincidence with the GRB (grey, dashed/red line). 
The flux of SN\,1998bw has been rescaled to match that of SN\,2006aj at the 
peak of the [\OI] line. 
{\em See the electronic paper for a color version of this figure.}
\notetoeditor{solid and grey lines refer to the b/w, printed version;
blue and red to the electronic, colour figure (f2_col.eps)}
}
\end{figure}


\begin{figure}
\plotone{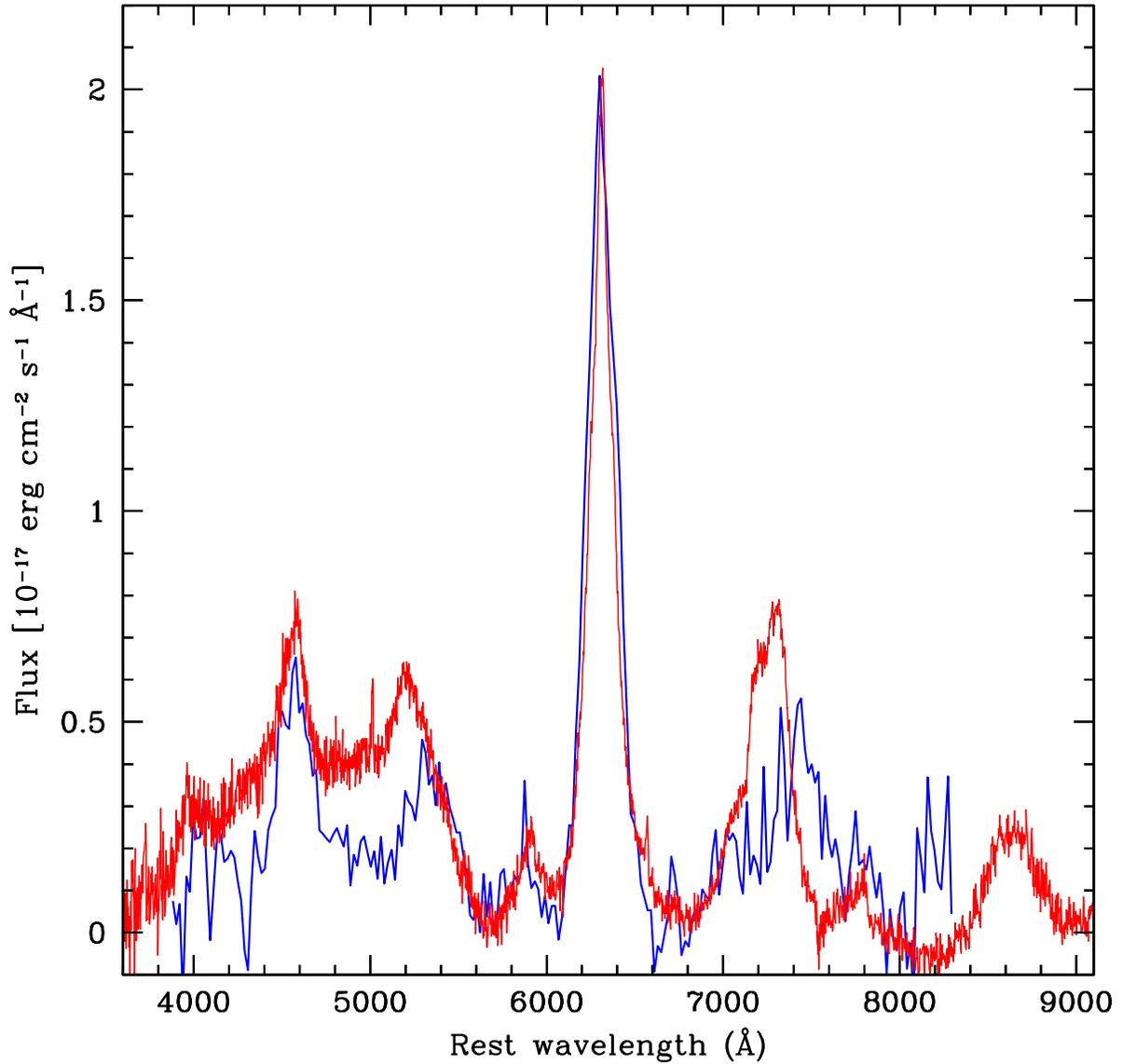}
\figcaption[]{A comparison of the VLT 2006 Sept 19--20  
(206 rest-frame days after XRF\,060218) spectrum of SN\,2006aj, calibrated and
binned with a 20~\AA\ boxcar (solid/blue line), 
with the SN\,1998bw spectrum obtained on 1998 November 26 (Patat et al. 2001), 
214 rest-frame days after GRB\,980425 (grey, dashed/red line).
{\em See the electronic paper for a color version of this figure.}
\notetoeditor{solid and grey lines refer to the b/w, printed version;
blue and red to the electronic, colour figure (f3_col.eps)}
}
\end{figure}


\begin{figure}
\plotone{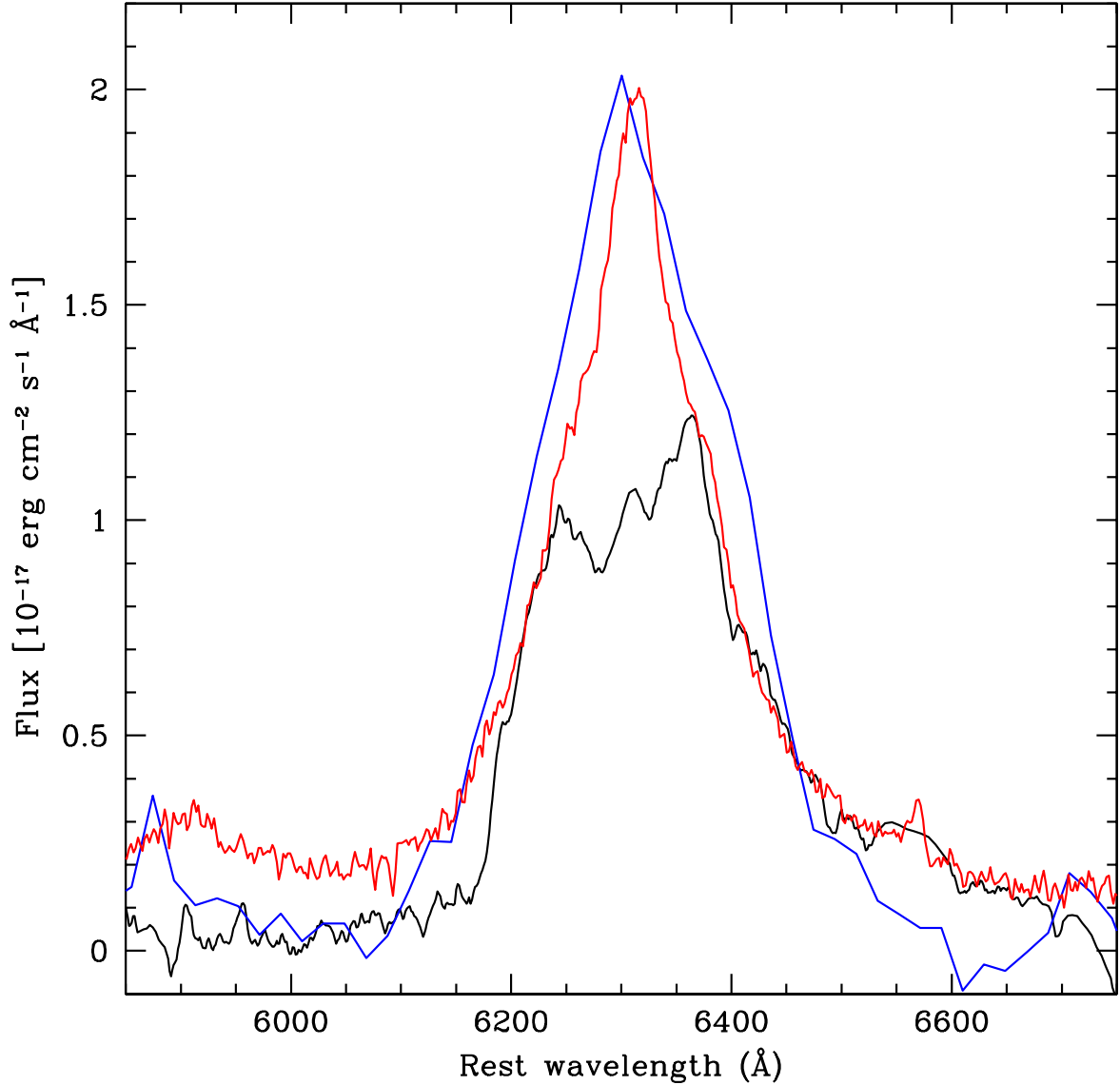}
\figcaption[]{A comparison of the [\OI] $\lambda\lambda$6300, 6363 line of 
SNe 2006aj (solid/blue line), 1998bw \citep[dotted/red line,][]{patat01}, and 
2003jd \citep[dashed/black line,][]{maz05}. The profile of the line in 
SN~2006aj is less peaked than that of SN~1998bw, although the average expansion 
velocity of other elements is larger in SN~1998bw than in SN~2006aj, 
indicating a smaller degree of asphericity.
{\em See the electronic paper for a color version of this figure.}
\notetoeditor{solid, dotted and dashed lines refer to the b/w, printed version;
blue, black, and red to the electronic, colour figure (f4_col.eps)}
}
\end{figure}


\begin{figure}
\plotone{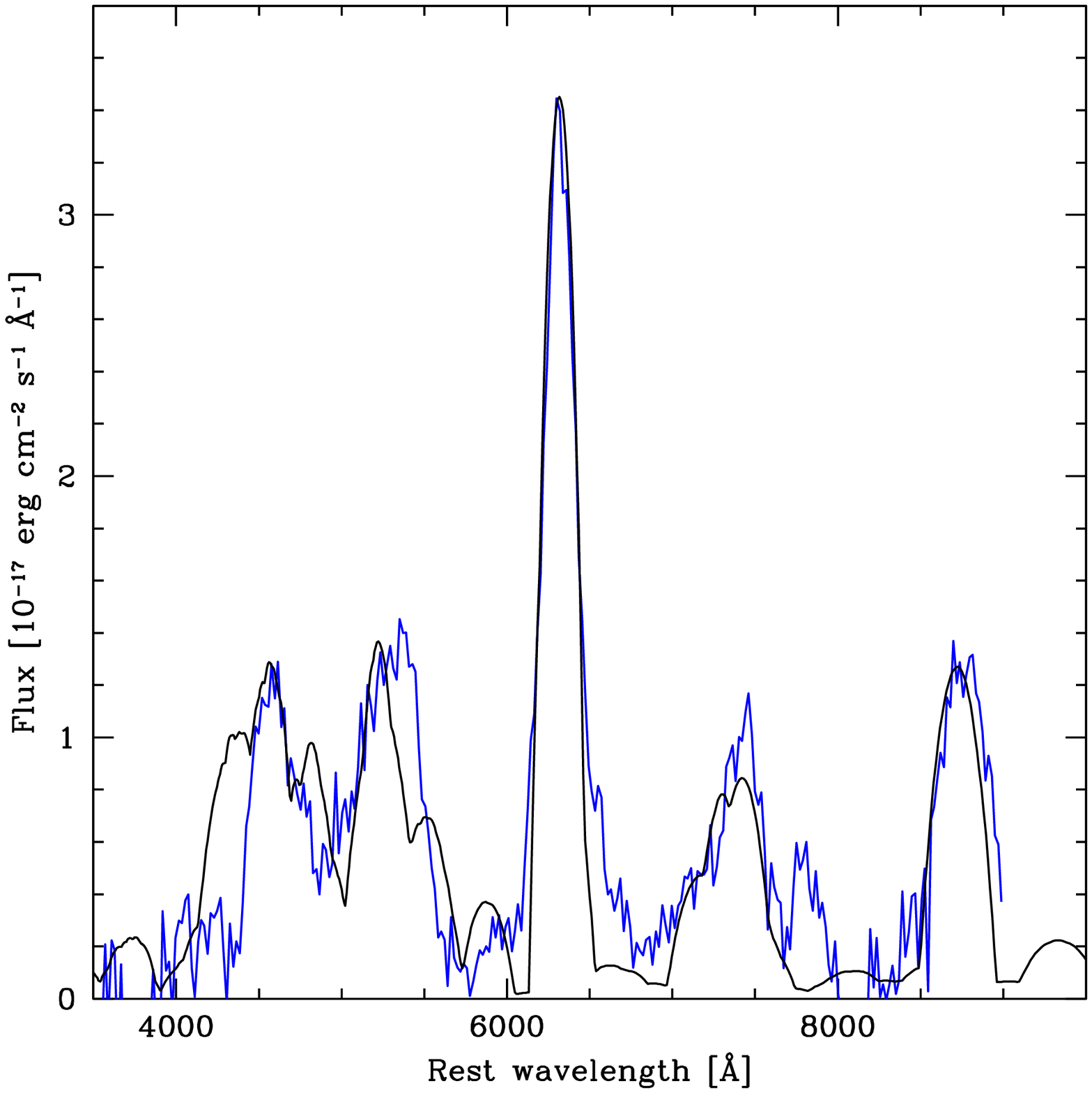}
\figcaption[]{The Keck spectrum of 2006 July 26 (grey/blue line) 
and the corresponding synthetic spectrum (black line).
{\em See the electronic paper for a color version of this figure.}
\notetoeditor{grey lines refer to the b/w, printed version;
blue lines to the electronic, colour figure (f5_col.eps)}
}
\end{figure}


\begin{figure}
\plotone{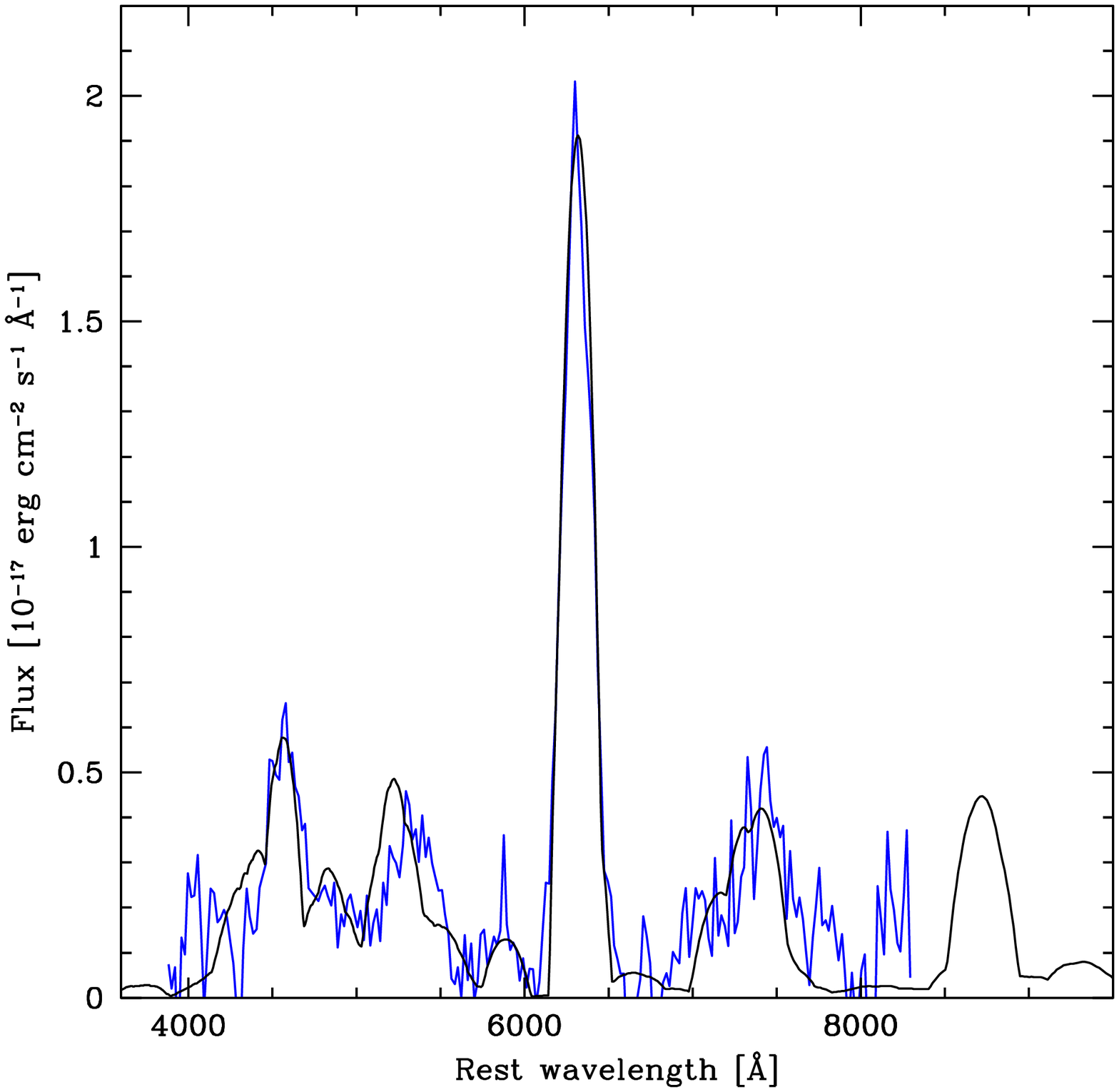}
\figcaption[]{The VLT spectrum of 2006 September 19--20 (grey/blue line) 
and the corresponding synthetic spectrum (black line).
{\em See the electronic paper for a color version of this figure.}
\notetoeditor{grey lines refer to the b/w, printed version;
blue lines to the electronic, colour figure (f6_col.eps)}
}
\end{figure}


\clearpage

\begin{deluxetable}{cccccccc}
\tabletypesize{\scriptsize}
\tablecaption{Model Properties}
\tablewidth{0pt}
\tablehead{
\colhead{UT Date} & 
\colhead{Epoch} &
\colhead{$v$} &
\colhead{$M$(\Nifs)} &
\colhead{$M_{\rm ej}$} &
\colhead{$M$(O)} &
\colhead{$M$(C)} &
\colhead{$L$} \\
\colhead{} &
\colhead{(rest-frame days)} &
\colhead{(\kms)} &
\colhead{($\Msun$)} &
\colhead{($\Msun$)} &
\colhead{($\Msun$)} &
\colhead{($\Msun$)} &
\colhead{(erg s$^{-1}$)} 
}
\startdata
2006 July 26 & 153 & 8000 & $0.20\pm0.01$ & $2.07\pm0.20$ & $1.50\pm0.15$ & $0.30\pm0.05$ & $1.3 \times 10^{41}$ \\
2006 Sept 19--20 & 206 & 7400 & $0.19\pm0.01$ & $1.94\pm0.25$ & $1.42\pm0.15$ & $0.25\pm0.10$ & $5.1 \times 10^{40}$ \\
\enddata
\end{deluxetable}


\end{document}